\documentclass{aa}

\usepackage{graphicx}
\usepackage{txfonts}
\usepackage{xcolor}
\usepackage{soul}
\newcommand{\LAT}{LAT flaring state}
\newcommand{\PCA}{PCA flaring state}
\newcommand{\CGRO}{CGRO dataset}
\newcommand{\eps}{\epsilon}
\newcommand{\Metsa}{Mets\"ahovi}
\newcommand{\revi}[1]{#1}

\begin{document}

\title{The high energy spectrum of 3C 273}
\subtitle{}
\author{V. Esposito,\inst{1,2}
          R. Walter,\inst{1,2}
          P. Jean,\inst{3,4}
          A. Tramacere,\inst{1}
          M. T\"urler,\inst{1,2}
          A. L\"ahteenm\"aki\inst{5,6}
          M. Tornikoski\inst{5}
          }

   \institute{Department of Astronomy, University of Geneva, ch. d'Ecogia 16, 1290 Versoix, Switzerland\\
              \email{Valentino.Esposito@unige.ch}
         \and
	      Geneva Observatory, University of Geneva, ch. des Maillettes 51, 1290 Versoix, Switzerland
	 \and
              Université de Toulouse; UPS-OMP; IRAP;  Toulouse, France
         \and
              CNRS; IRAP; 9 Av. colonel Roche, BP 44346 F-31028 Toulouse cedex 4, France
         \and
              Aalto University \Metsa\ Radio Observatory, \Metsa ntie 114, FIN-02540 Kylm\"al\"a, Finland
         \and
              Aalto University Department of Radio Science and Engineering,  P.O. BOX 13000, FI-00076 Aalto, Finland.
             }

   \date{}
 

  \abstract
   {}
   {The high energy spectrum of 3C 273 is usually understood in terms of inverse-Compton emission in a relativistic leptonic jet. This model predicts variability patterns and delays that could be tested with simultaneous observations from the radio to the GeV range.}
   {The instruments IBIS, SPI, JEM-X on board INTEGRAL, PCA on board RXTE, and LAT on board Fermi have enough sensitivity to follow the spectral variability of 3C 273 from the keV to the GeV. We looked for correlations between the different energy bands, including radio data at 37 GHz collected at the \Metsa\ Radio Observatory and built quasi-simultaneous multiwavelength spectra in the high energy domain when the source is flaring either in the X-rays or in the $\gamma$ rays.}
   {Both temporal and spectral analysis suggest a two-component model to explain the complete high energy spectrum. X-ray emission is likely dominated by a Seyfert-like component while the $\gamma$-ray emission is dominated by a blazar-like component produced by the relativistic jet. The variability of the blazar-like component is discussed, comparing the spectral parameters in the two different spectral states. Changes of the electron Lorentz factor are found to be the most likely source of the observed variability.}
   {}


   \maketitle

\section{Introduction}
\label{intro}
Active galactic nuclei (AGN) are very energetic extragalactic sources powered by accretion on supermassive black holes. Blazars are a class of AGN characterized by a relativistic jet pointing towards the observer. Because of the beaming, the emission originating in the jet dominates the observed spectrum.

The quasar 3C 273 is the brightest and hence one of the best monitored AGN. Located at a distance of $\mathrm{z} = 0.158$ \citep{Strauss1992}, it features interesting blazar-like (a jet with superluminal motion and high variability) and Seyfert-like (the strong blue bump, the soft X-ray excess, and variable emission lines) characteristics.

Several observational campaigns have been carried out on 3C 273, starting from the broad multiwavelength campaign by \citet{Lichti1995}, which made use of data spanning from radio to $\gamma$ rays, to more recent observations in X-rays. \citet{Courvoisier2003} described simultaneous observation of INTEGRAL, XMM-Newton, and RXTE performed in January 2003. \citet{Chernyakova2007} used simultaneous INTEGRAL and XMM-Newton data in the period 2003--2005 and compared their results to historical data to investigate the secular evolution of 3C 273. \citet{Liuliu2011} used XMM-Newton data to investigate the nature of the soft X-ray excess.

\begin{figure*}
\centering
\includegraphics[angle=-90, width=\hsize]{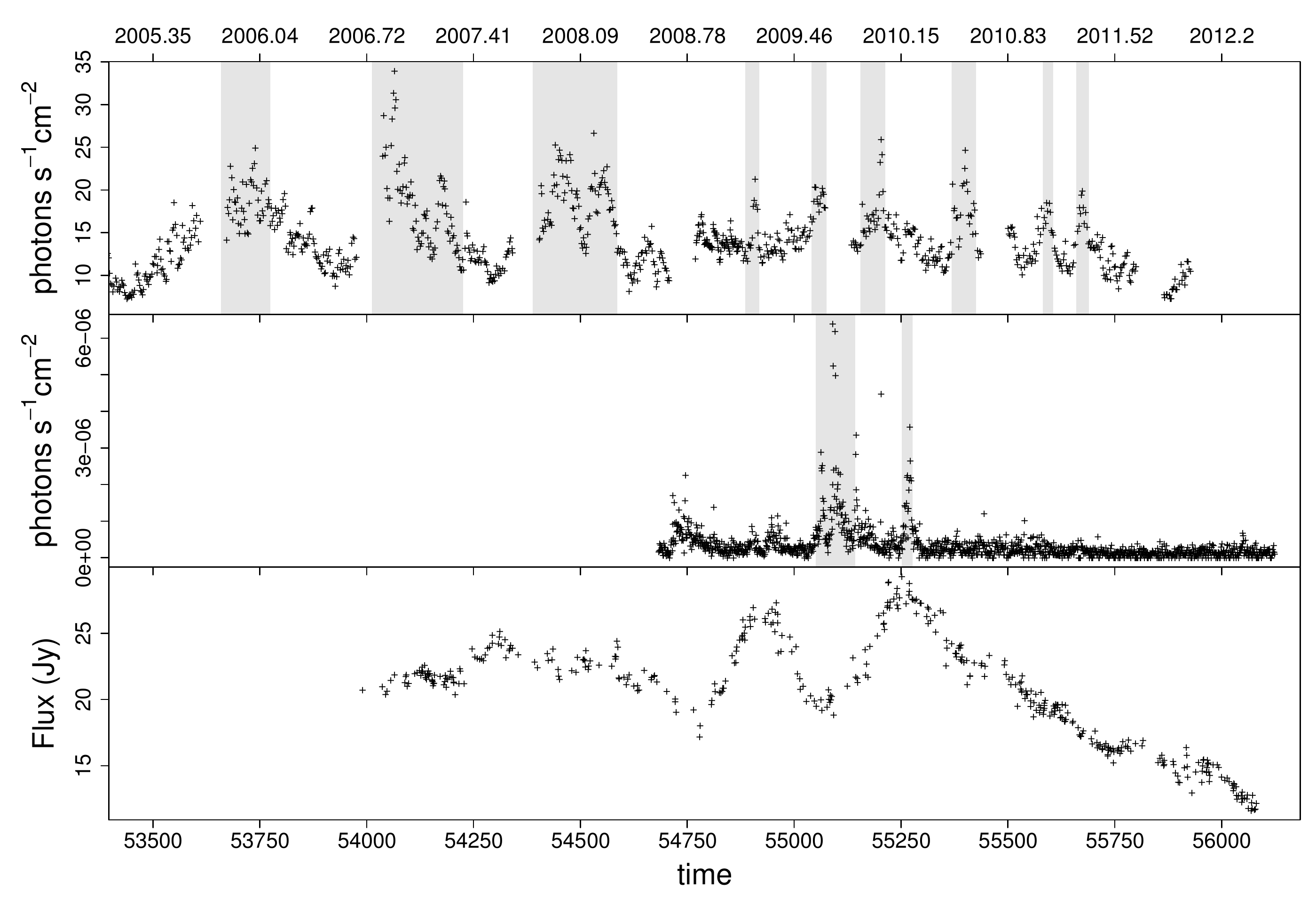}
\caption{\textit{Top panel:} RXTE-PCA lightcurve in 0.2 - 75 keV. \textit{Middle panel:} Fermi-LAT lightcurve in 0.1 - 100 GeV. Area marked in light grey colour shows the selected flaring time epochs in both lightcurves. \textit{Bottom panel:} radio lightcurve at 37 GHz. Time units are days on bottom axis (Modified Julian Date) and decimal years on top axis.}
\label{fig_fltime}
\end{figure*}

Many works have been published to describe in detail the spectral and temporal characteristics of 3C 273, which has been widely used as a test case to validate theoretical models and predictions. \citet{Paltani1998} found two distinct variable components in the optical--ultraviolet band. \citet{Soldi2008} studied the temporal variability in a wide energy range. \citet{Grandi2004} were able to decouple the non-thermal (jet) and thermal (accretion flow) emission by studying the long-term spectral variability of BeppoSAX data. \citet{Marcowith1995} tested their model of relativistic electron--positron beam for blazar emission. \citet{Marscher1985} and \citet{Turler2000} modelled the radio emission in the context of synchrotron emission from relativistic electrons produced by shocks in the jet. \citet{Artyukh2012} investigated the physics and the condition of the core of 3C 273 on parsec scale. A detailed review has been written by \citet{Courvoisier1998} and a database collecting public data of the source over several decades is available online \citep{Turler1999b,Soldi2008}.

The aim of this work is to use contemporaneous X-ray to $\gamma$-ray observations obtained with INTEGRAL and Fermi for the first time since the CGRO campaign \citep{Lichti1995} to study the origin of the non-thermal spectral components. In Sect. \ref{data} we introduce the data used for this work and explain how the datasets have been buildt. In Sect. \ref{temp} we present the results of the cross-correlation analysis between lightcurves obtained with Fermi-LAT, RXTE-PCA, and at 37 GHz and compare our results with those of previous studies. We find that $\gamma$-ray and radio emission are correlated and that the X-ray component has a different origin.

In Sect. \ref{spec} we describe the high energy spectrum of 3C 273 and its variability. Data from RXTE-PCA (0.2 - 75 keV), JEM-X, ISGRI and SPI on board INTEGRAL (3 keV - 1 MeV), and Fermi-LAT (100 MeV - 100 GeV) are used. We selected two different spectral states based on the behaviour of the PCA and LAT lightcurves, and fit several models to their spectra to investigate the differences between the two states. We find that the best fit model is a two-component model, which can be interpreted as a Seyfert-like component dominating the X-ray emission, and blazar-like component dominating at $\gamma$ rays. In Sect. \ref{disc} we discuss the results and give a physical interpretation of the data, focusing on the blazar-like component, in the context of relativistic jet emission and synchrotron self Compton (SSC) model, and finally we summarize the results in sect. \ref{concl}.

\section{The data sample}
\label{data}
\revi{For this work we used data from RXTE-PCA, INTEGRAL and Fermi-LAT, which provide a broad spectral coverage from a few keV up to 100 GeV, and radio data collected at the \Metsa\ Radio Observatory\footnote{http://metsahovi.aalto.fi/en/}.}

\revi{Fermi-LAT \citep[0.1 - 100 GeV,][]{Atwood2009} data are used to study temporal and spectral variability at high energies. Thanks to its large field of view, Fermi-LAT has observed 3C 273 almost continuously. The SED is extracted with the Maximum Likelihood fitting technique commonly used to analyse Fermi data and the light curve is built using the LAT Aperture Photometric method assuming the average spectral index of 2.45 as listed in the Fermi-LAT 2nd Catalog \citep{Nolan2012} and a binning time of 1 day. The data spans from August 2008 to July 2012 and they are extracted using the Fermi Science Tools version v9r27p1 and the instrument response function P7\_V6.}

\revi{INTEGRAL \citep{Winkler2003} JEM-X (6 - 20 keV), ISGRI (17 - 210 keV) and SPI (50 - 770 keV) data are used for the spectral analysis. The JEM-X and ISGRI data are analysed with OSA 10 \footnote{http://www.isdc.unige.ch/integral/analysis}, while SPI spectra are built with the SPI\_TOOLSLIB tool \citep{Knodlseder2004} and the spectra extracted by a model fitting method. They span from January 2003 to January 2012.}

\revi{RXTE-PCA \citep[0.2 - 75 keV,][]{Jahoda1996} data are collected with the HEAVENS web interface\footnote{http://www.isdc.unige.ch/heavens/} \citep{Walter2010} and span from March 2005 to December 2011. During this period 3C 273 has been observed by RXTE-PCA every few days for $ \sim $ 2 ks with only seven long gaps (see Fig. \ref{fig_fltime} upper panel).}

\revi{The radio data at 37 GHz are obtained at the \Metsa\ Radio Observatory and span from September 2006 to June 2012. They are used for timing analysis to probe the correlation with the high energy emission. The flux density scale is set by observations of DR 21. Sources NGC 7027, 3C 274 and 3C 84 are used as secondary calibrators. A detailed description of the data reduction and analysis is given in \citet{Terasanta1998}. The error estimate in the flux density includes the contribution from the measurement rms and the uncertainty of the absolute calibration.}

\section{Temporal Analysis} 
\label{temp}

\subsection{Cross-Correlation}
\label{temp_cc}

In the multiwavelength variability study of 3C 273 by \citet{Soldi2008}, correlations between the lightcurves obtained at various frequencies were studied. In particular hard X-rays ($\gtrsim 20~\mathrm{keV}$) and radio-mm were found to be correlated, while no correlation was found with lower energy X-rays ($\lesssim 20~\mathrm{keV}$) or radio-mm lightcurves. A difference between the hard ($\gtrsim 20~\mathrm{keV}$) and soft X-rays was also found in the fractional variability spectrum. This suggests a different origin for the X-rays above and below $\sim 20~\mathrm{keV}$.\\

In order to study the origin of the high energy emission we used lightcurves obtained with Fermi-LAT, RXTE-PCA and in the radio with the \Metsa\ Radio Telescope. 
The lightcurves are shown in Fig. \ref{fig_fltime}.

The LAT and PCA lightcurves show a different behaviour. The LAT data show a stable quiescent flux at $\sim 0.3 \cdot 10^{-6}~\mathrm{photons~s^{-1}~cm^{-2}}$ with small flares randomly distributed and three very strong flares in September 2009 (MJD from 55071 to 55097). These strong flares are described in detail by \citet{Abdo2010} and \citet{Rani2013}. \citet{Rani2013} found a variability time scale of the order of $\sim 0.1$ day for the shortest flare and of $\sim 0.5$ -- 1 day for the others. The PCA lightcurve shows several consecutive flares with alternating high and low states, the rate varying from $\sim 5$ to $\sim 30~\mathrm{photons~s^{-1}~cm^{-2}}$.\\

\begin{figure}
\centering
\includegraphics[width=\hsize]{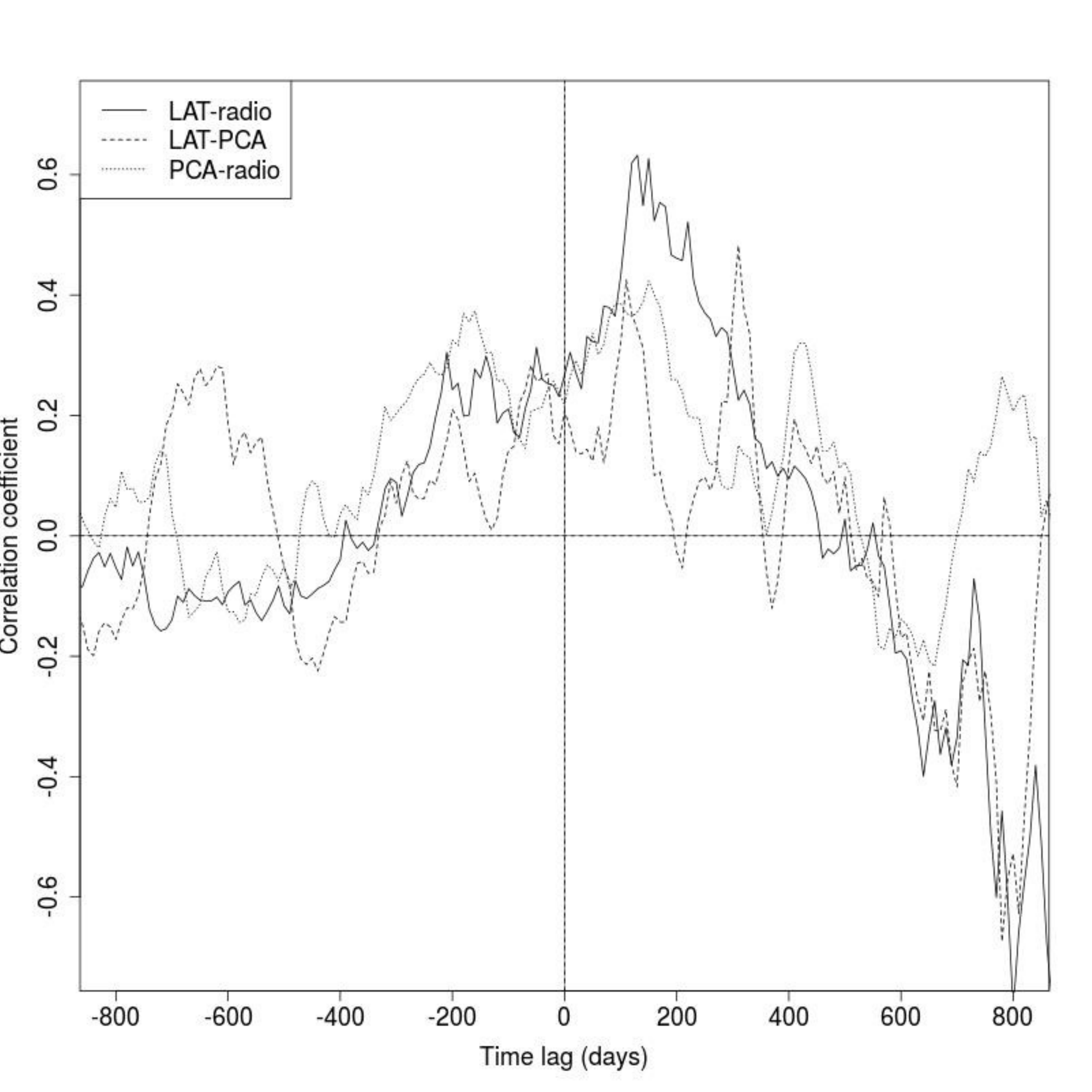}
\caption{Cross-correlation between Fermi-LAT, RXTE-PCA and radio lightcurves. A positive lag for an "X-Y" correlation means a delay of Y with respect to X.}
\label{fig_corr}
\end{figure}

To study correlations, we used the discrete correlation function \citep[DCF, ][]{Edelson1988} with a time bin of 10 days. In a cross-correlation between the "X" and "Y" lightcurves, labelled "X-Y" in Fig. \ref{fig_corr}, a positive time lag means that the Y lightcurve is delayed with respect to X. Cross-correlation curves are shown in Fig. \ref{fig_corr}. The solid line represents the cross-correlation between the radio and the LAT data. The correlation is peaked at $\sim +120$ days with a coefficient of $\sim 0.63$ suggesting that $\gamma$ rays are leading the radio emission. Dashed and dotted lines represent the absence of significant correlation of the PCA lightcurve with the LAT or radio lightcurves.

This result is in agreement with the observation of \citet{Soldi2008}, suggesting that the hard X-ray photons come from the jet, whereas Softer X-rays instead are dominated by a different component.

\subsection{Using the IC response}
\label{temp_conv}
\begin{figure}
\centering
\includegraphics[angle=-90, width=\hsize]{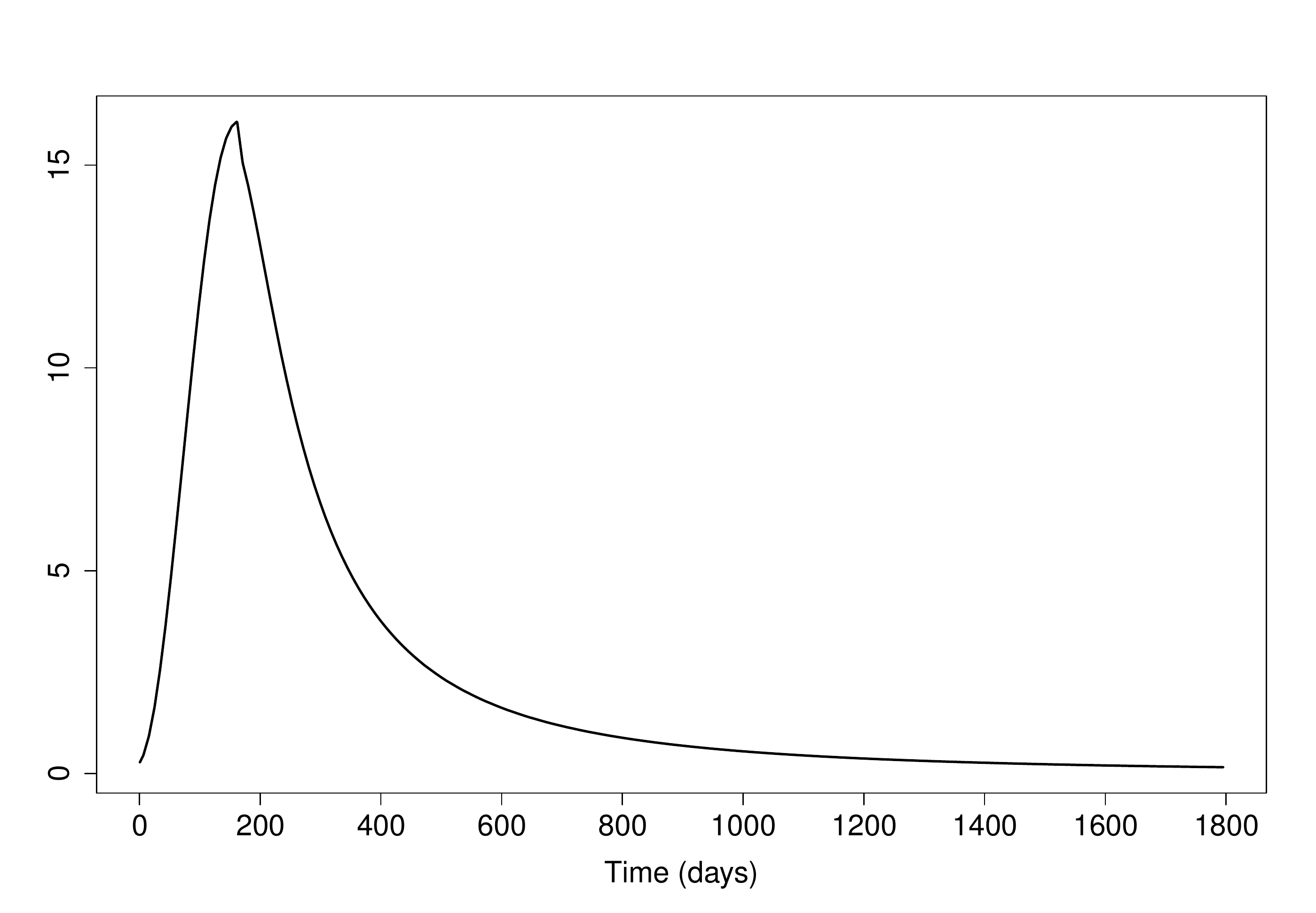}
\caption{Profile of the best radio response, the \textit{y} axis is in arbitrary scale. The maximum is located at 160 days after the beginning of the curve. The FWHM of the curve is 197 days. See text for details.}
\label{fig_resp}
\end{figure}

\begin{figure}
\centering
\includegraphics[angle=-90, width=\hsize]{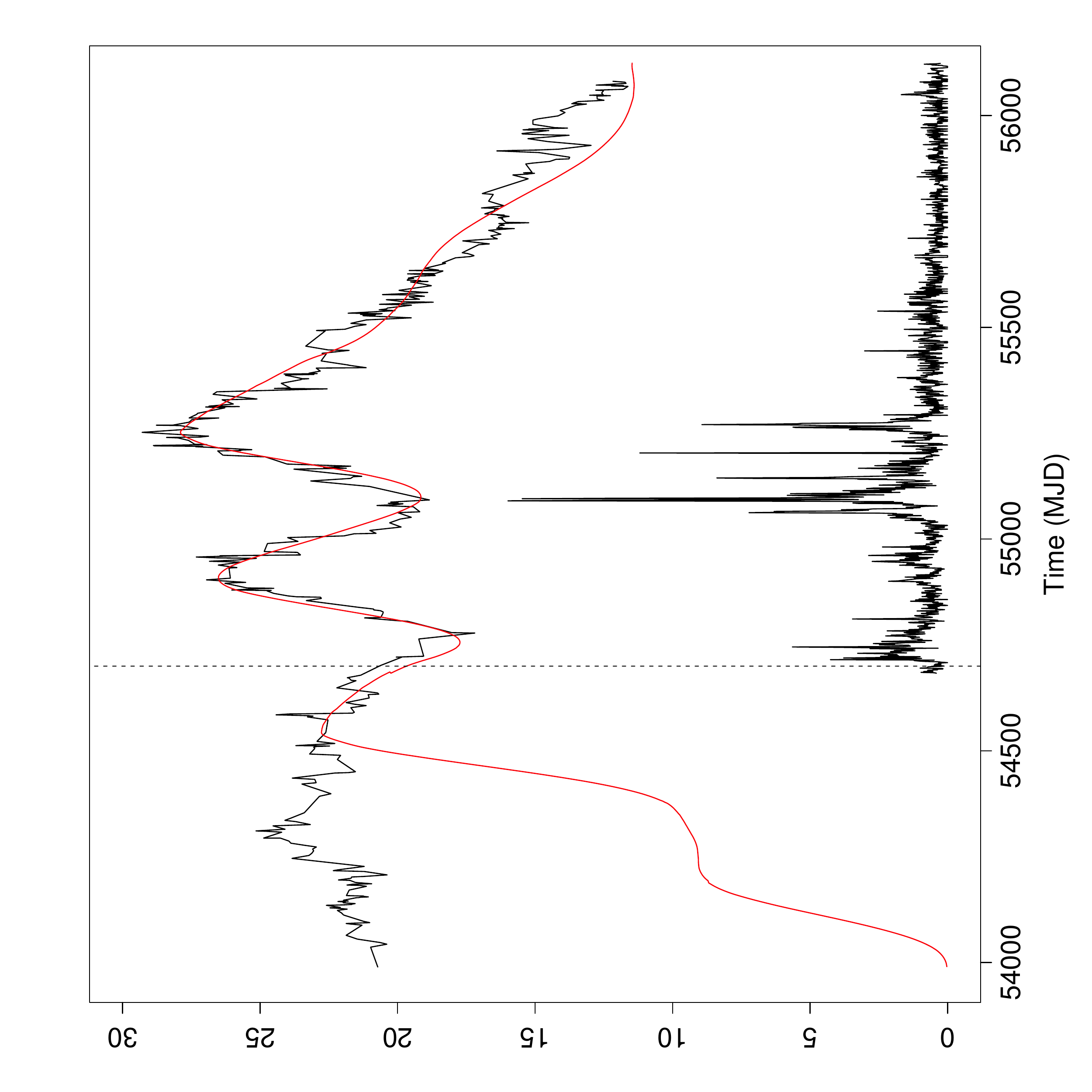}
\caption{Radio lightcurve (black) compared with the convolution obtained from the LAT lightcurve (red). Fermi-LAT lightcurve is shown in the bottom of the picture for comparison (not in scale). Vertical dotted line is MJD = 54700 days. See text for details.}
\label{fig_conv}
\end{figure}



The connection between the radio and the $\gamma$-ray emission in radio-loud AGN has been widely investigated proving that radio and $\gamma$-ray luminosities are correlated (see for example \citet{Padovani1993,Valtaoja1999,Jorstad2001,Lahteenmaki2003,Marscher2008,Hovatta2009,Ackermann2011}).

In order to better study the link between the GeV and the radio emission, we try to evaluate whether the observed radio variability could arise from a simple convolution of the Fermi lightcurve. 
The response function would convolve each spike of the LAT lightcurve to a broader and delayed radio flare. The sum of all these broadened flares, which may be superimposed, should match the observed radio lightcurve. The response would then naturally introduce the delay seen with the cross-correlation of Fig. \ref{fig_corr}, caused by the cooling time of the electrons emitting synchrotron radiation and the expansion of the emitting region, becoming optically thin at lower energies.

As the convolution function is not known a priori, we rely on the shape of a typical outburst in 3C 273 at radio frequencies. \citet{Turler1999,Turler2000} developed a method -- in the frame of a shock-in-jet scenario -- to extract from the long-term multi-wavelength lightcurves the temporal evolution of a typical synchrotron outburst at any frequency. We use here the model lightcurve of an average outburst of 3C 273 at 37 GHz as derived from the observations obtained between 1980 and 2000 and with the approach and parameters described by \citet{Turler2007}. This model lightcurve rises relatively steeply to reach a maximum in less than a year before decaying gradually.\\

With this response function we could not reproduce the radio data because the peak of the response profile is too late and broad when compared to the observations. The observed radio lightcurve can only be reproduced by compressing the response by a factor of 2 in time scale and shifting it by $-30$ d. This compressed and shifted response, shown in Fig. \ref{fig_resp}, allow us to match the width and the position of the observed radio flares.

The necessity to shrink the response can be explained considering that the convolution smears the complex $\gamma$-ray variability, with the result that a sequence of short $\gamma$-ray flares separated by short time lag will be observed in the radio as a single broad flare. As explained above, the response curve was constructed on long-term multi-wavelength lightcurves, which do not include $\gamma$-ray data. It is hence possible that what was seen in the construction of the response as a single flare, i.e. a single shock wave in the modelling of \citet{Turler1999,Turler2000}, might actually be the superposition of several smaller shocks.\\

It has to be pointed out that the available LAT data start later than the radio data. So we cannot reproduce correctly the early radio signal. We added therefore some fake $\gamma$-ray data to reproduce the radio signal at MJD $\sim$ 54700.

We also needed to change the amplitude of the response from flare to flare to match correctly the two peaks in the radio band at MJD $\sim 55000$ and $\sim 55400$. To obtain a qualitatively good match between our synthetic light curve and the radio data we used four different response normalizations. There is no physical constraint for these responses to be the same for different radio outbursts: the physical details of the shocks are inside the response function, and different shocks may undergo different physical conditions. Table \ref{tab_pconv} shows the scaling factors used to modify the original response and the time intervals. The first response is used to build the synthetic lightcurve before the beginning of LAT data (relying on the fake $\gamma$-ray data introduced). The scaling factor is arbitrary chosen to be relative to the amplitude of the response in the first time interval.

With these prescription we are able to qualitatively reproduce the radio light curve after MJD $\sim 54700$. Figure \ref{fig_conv} shows the radio lightcurve (black line) and the curve derived from the LAT lightcurve with the described technique (red line). The LAT lightcurve is also drawn for comparison.
Since the maximum of the response is at $+160$ days (Fig. \ref{fig_resp}), the average lag between $\gamma$-ray and radio flares is therefore of the order of $140 \pm 20$ days.

\begin{table}
\caption{Time start and relative amplitude scaling of the several response profiles used to construct the synthetic lightcurve (red line in Fig. \ref{fig_conv}). See text for the details.}
\label{tab_pconv}
$$
\begin{array}{cc}
\hline\hline
\noalign{\smallskip}
\mathrm{Time\: interval\: in\: MJD} & \mathrm{Scaling\: factor} \\
\hline
\noalign{\smallskip}
53971 - 54710 & 1 \\
\hline
\noalign{\smallskip}
54710 - 54890 & 2.20 \\
\hline
\noalign{\smallskip}
54890 - 55340 & 1.04 \\
\hline
\noalign{\smallskip}
\mathrm{after}\:\: 55340 & 1.83 \\
\hline
\end{array}
$$
\end{table}

\section{Spectral Analysis}
\label{spec}
To differentiate and further investigate the nature of the X-rays and the hard X-ray to $\gamma$-ray components suggested by the cross-correlation analysis (Sect. \ref{temp}) we built the SED of 3C 273 in two different spectral states. We defined the \emph{PCA flaring state} as the time intervals when the source flux \revi{in the RXTE-PCA lightcurve} exceeds $18~\mathrm{photons~s^{-1}~cm^{-2}}$ and remain larger than this value for at least 9 days. Similarly we define the \emph{LAT flaring state} as the time intervals when the source flux \revi{of the Fermi-LAT lightcurve} goes beyond $1.2 \cdot 10^{-6}~\mathrm{photons~s^{-1}~cm^{-2}}$ and remain larger than this value for at least 3 days. In case of two consecutive flares we expanded the selected interval to the full width of the flaring period. The epochs defining the flaring states are shadowed in Fig. \ref{fig_fltime}.

For both spectral states, we built a multiwavelength spectrum from the X-rays to the $\gamma$ rays, using data from RXTE, INTEGRAL, and Fermi selected in the time intervals identified as the PCA and \LAT s.


\revi{The signal-to-noise ratios are enough to build multiwavelength spectra for each of the datasets, but not to follow the temporal evolution of the spectral parameters during the individual flares. Moreover, INTEGRAL data are not available for the total duration of the flaring times identified.}

In the following discussion we will also compare our datasets with the SED obtained from the multiwavelength campaign on 3C 273 performed in 1991 \citep{Lichti1995} and built with data from GINGA (2 - 30 keV) and data from OSSE, COMPTEL and EGRET on board CGRO (20 keV - 10 GeV). X-ray and $\gamma$-ray data showed a break at $\sim 1~\mathrm{MeV}$. \citet{Lichti1995} tested several models on their whole SED which includes also radio, infrared, optical and ultraviolet data.

We first tested a simple cutoff powerlaw model $N(E) = N_{cp} E^{-\Gamma} \exp(-E/E_c)$, which is not able to fit the data: the $\chi_{red}^2$ is $\sim 48$ for the \LAT, $\sim 37$ for the \PCA\ and $\sim 4$ for the \CGRO. In the LAT and \PCA s the flux and the curvature of the LAT data is clearly not consistent with an exponential cutoff, and also in the \CGRO\ EGRET data are clearly above the exponential cutoff.

We then used several models to fit the datasets: a \textit{broken powerlaw}, a \textit{cutoff powerlaw} and a \textit{log-parabola}. None of these models has been used in \citet{Lichti1995}, where the high energy data were fitted with an empirical function to model the break at $\sim 1~\mathrm{MeV}$.

The log-parabola model (see sect. \ref{spec_lpmodel}) is able to reproduce the typical inverse-Compton bumps measured by Fermi in blazar spectra, which show an intrinsic curvature \citep{Nolan2012} and can be interpreted with stochastic acceleration of the electrons \citep[and reference therein]{Massaro2004,Tramacere2011}.\\

\begin{figure}
\centering
\includegraphics[angle=-90, width=\hsize]{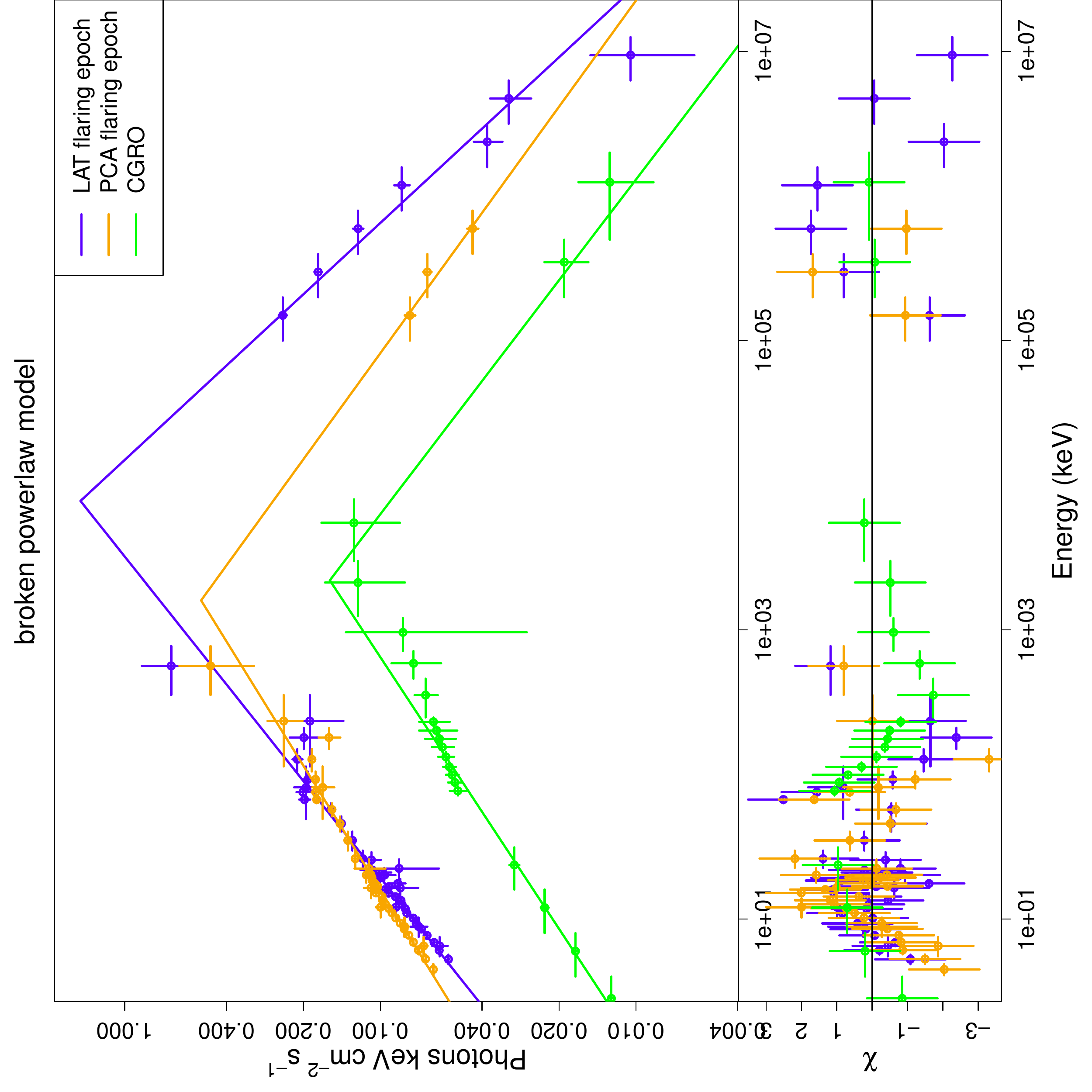}
\caption{3C 273 multiwavelength spectra at different epochs fitted with the broken powerlaw model. The upper panel shows the SED and the best fit models (continuous lines), the lower panel shows the residuals from the best fit models.}
\label{fig_bkp}
\end{figure}
\begin{figure}
\centering
\includegraphics[angle=-90, width=\hsize]{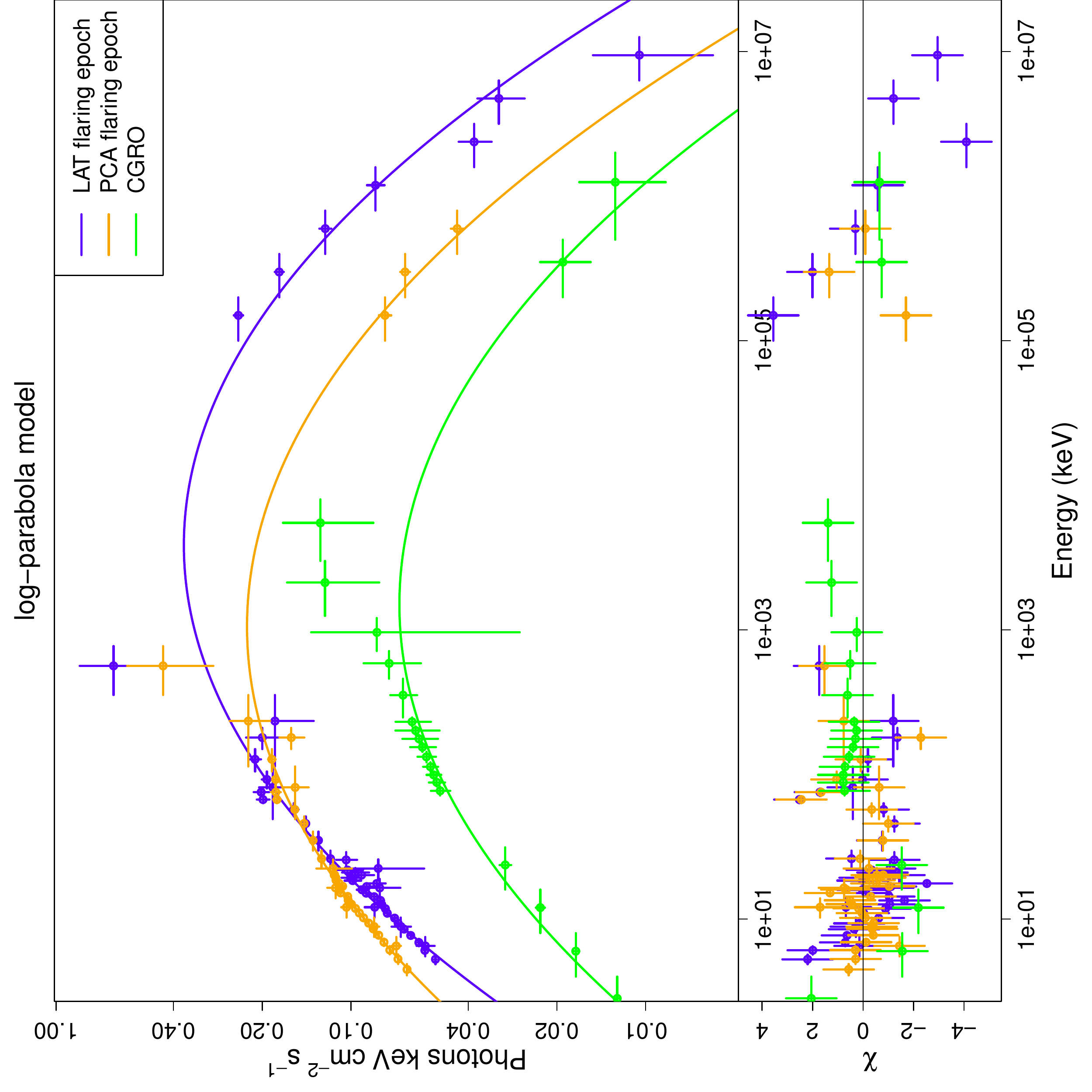}
\caption{Same as Fig. \ref{fig_bkp} for the log-parabola model.}
\label{fig_lp}
\end{figure}
\begin{figure}
\centering
\includegraphics[angle=-90, width=\hsize]{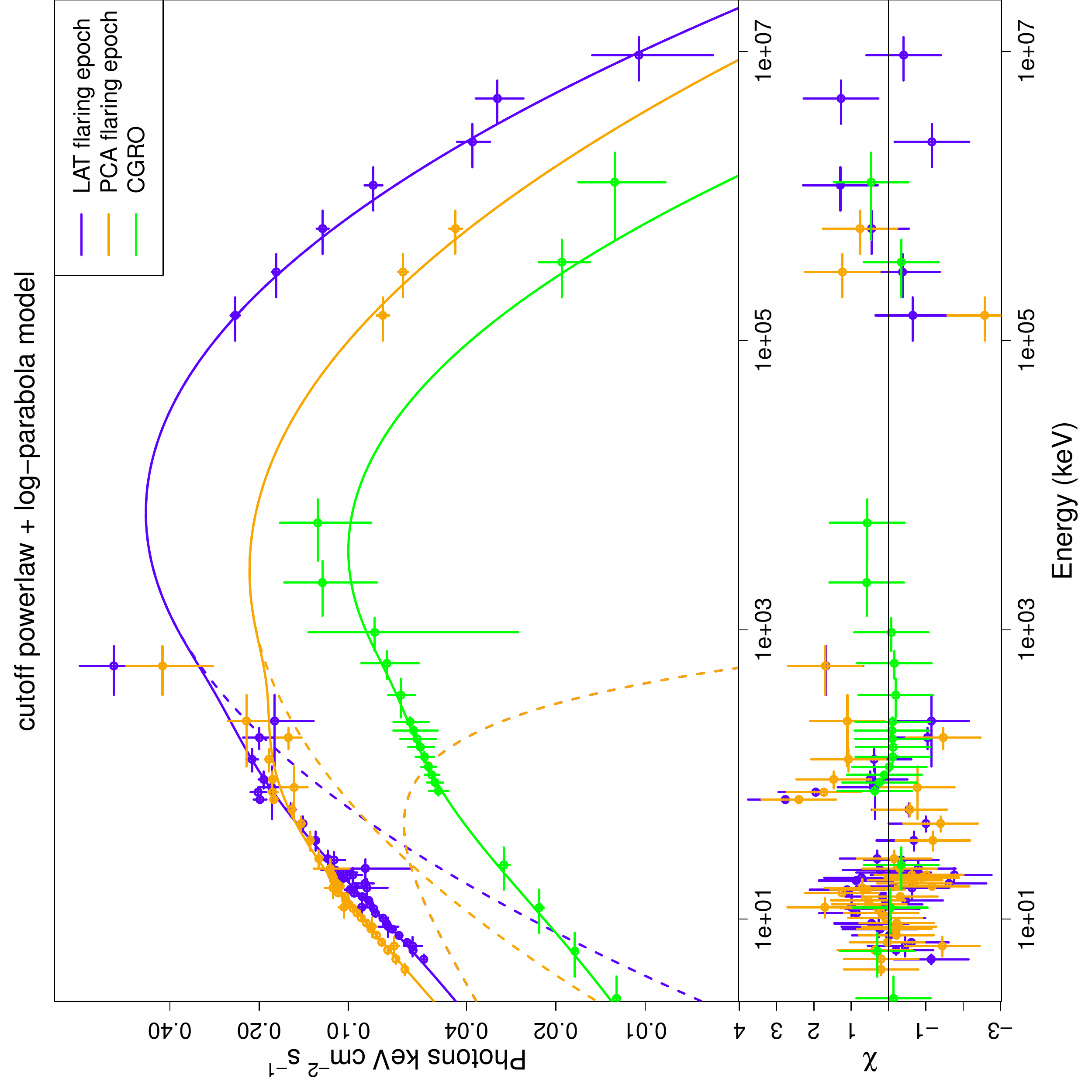}
\caption{Same as Fig. \ref{fig_bkp} for the cutoff powerlaw + log-parabola model. The dashed lines shows the two components of the model for the LAT and \PCA s.}
\label{fig_lpcop}
\end{figure}

Figures \ref{fig_bkp}, \ref{fig_lp} and \ref{fig_lpcop} show the multiwavelength spectra for the LAT and \PCA s, plus the \CGRO\ from \citet{Lichti1995}. Tables \ref{tab_bkp}, \ref{tab_lp} and \ref{tab_lpcop} list the best fit parameters, reduced $\chi^2$ and degree of freedom for each dataset and each model tested. Errors are at $2.7 \: \sigma$ and the parameters without error are fixed.

In general the \CGRO\ is closer to the \PCA\ whatever model is used to fit the data. The main difference is the overall luminosity, which is systematically lower than during the LAT and \PCA s. \revi{We have also built the spectrum of the quiescent state, i.e. when the source is not flaring either in RXTE-PCA or in Fermi-LAT. It has a similar shape and is 2 times brighter than observed by CGRO.}

\subsection{Broken powerlaw model}
\label{spec_bkpmodel}
Fig. \ref{fig_bkp} shows the three datasets with the best fit for a broken powerlaw model:

\begin{equation*}
\label{eq_bknpow}
N(E) = \begin{cases} N_{bp}~E^{-\Gamma_1} & \mbox{ at } E < E_b\\ N_{bp}~E_b^{\Gamma_2 - \Gamma_1} E^{-\Gamma_2} & \mbox{ at } E > E_b \end{cases}
\end{equation*}

The best fit parameters are reported in Table \ref{tab_bkp}. The fit by means of the broken powerlaw model does not provide a good description of the LAT and \PCA s spectra, but the variation of the spectral indices between the two datasets gives some hints on the physics. This difference leads to a variation of the spectral break $\Delta \Gamma = \Gamma_2 - \Gamma_1$ from $\sim 0.7 \pm 0.1$ for the \PCA\ to $\sim 1.0 \pm 0.05$ for the \LAT, which could be related to a variation of the average slopes of the electron distribution (assuming the electron distribution modelled with a broken powerlaw too) from the LAT to the \PCA. The break energy $E_b$ is larger for the \LAT\ than for the \PCA.

Both datasets show a bump at $\sim 20$ keV (in RXTE and JEMX data), clearly visible in the lower panel of Fig. \ref{fig_bkp}: to understand if this structure is real, we built the spectrum of the Crab nebula to compare the deviations from the best fit model for the two sources. As the deviations are of the same order, we conclude that this bump is likely due to a calibration issue.

\subsection{Log-parabola model}
\label{spec_lpmodel}
The log-parabola model is given by

\begin{equation}
\label{eq_logpar}
N(E) = N_{lp} \left( \frac{E}{E_{0}} \right) ^ {-\alpha - \beta \log \left( \frac{E}{E_{0}} \right)},
\end{equation}

\noindent where $N_{lp}$ is the photon flux at $E = E_{0}$, the $\alpha$ index denotes the slope at $E = E_{0}$ and the $\beta$ index models the curvature of the log-parabola. 

Figure \ref{fig_lp} shows the three datasets with the best fit obtained with the log-parabola model. The best fit parameters are reported in Table \ref{tab_lp}. For the \LAT\ the log-parabola alone is not able to fit the data adequately ($\chi_{red}^2 > 2$). The residuals indicate that the intrinsic curvature of the LAT data does not correspond to that of the complete SED.

The log-parabola model provides a good fit for the \PCA, but it has to be pointed out that the LAT data in this case have worse statistics than for the \LAT. Consisting of 3 points only, they do not show a clear intrinsic curvature. In this case the $\beta$ parameter can be determined only from the multiwavelength fit and it is impossible to state if an extra component is needed or not. A similar statement can be done for the \CGRO, where EGRET data do not have enough statistics to accurately constrain the shape of the spectrum at $\gamma$-ray energies.

\subsection{Cutoff powerlaw + log-parabola model}
\label{spec_lpcopmodel}
From the cross-correlation analysis (Sect. \ref{temp_cc}) we inferred that X-rays and $\gamma$ rays are dominated by different components. Interpreting the LAT data emitted by inverse-Compton processes in the jet, we tried to add an extra component in the form of a cutoff powerlaw at lower energies to model the X-ray emission in a similar way as assumed in Seyfert galaxies.

Figure \ref{fig_lpcop} shows the three datasets with the best fit for the cutoff powerlaw + log-parabola model. The best fit parameters are reported in Table \ref{tab_lpcop}. Adding the cutoff powerlaw component improves the quality of the fit for the \LAT, supporting the hypothesis of an extra X-ray component. The parameters of the cutoff powerlaw component have been fixed to the typical values for Seyfert Galaxies \citep{Mushotzky1984,Turner1989} and we used the same values of spectral index and cutoff energy for the three datasets because fitting the datasets with six free parameters leads to a degeneracy, especially for the cutoff energy $E_c$. This model provides a good fit also for the \PCA\ and for the \CGRO.\\

In order to constrain the best value for the cutoff powerlaw component we also tried to fit the X-ray data only (PCA, JEM-X, ISGRI and SPI) for the \LAT\ with the cutoff powerlaw + log-parabola model. In this case, the parameters of the log-parabola component are fixed to the values that provide the best fit of LAT-Fermi data alone. The idea is to model the contribution of the blazar-like component fitting the $\gamma$-ray data alone and extrapolating its contribution at X-rays. Fitting the two components independently allows us to derive the best fit parameters for the Seyfert-like component, modelled as a cutoff powerlaw (Table \ref{tab_lpcop2}). It is not possible to follow the same procedure for the \PCA\ due to lack of statistics of the LAT data.\\

We can also fit the whole multiwavelength spectrum using the spectral parameters of Table \ref{tab_lpcop2} for the \LAT\ instead of the fixed parameters previously chosen (Table \ref{tab_lpcop}). In this way we obtain a new set of best-fit parameters and reduced $\chi^2$ for the cutoff powerlaw + log-parabola model (Table \ref{tab_lpcop3}). As expected the $\chi^2_{red}$ is better for the \LAT, but it is worse for the \PCA, suggesting that the cutoff powerlaw component may not have the same parameters in both datasets. The last line of Table \ref{tab_lpcop3} reports the best fit obtained leaving the cutoff powerlaw photon index $\Gamma$ free to vary.

\begin{figure}
\centering
\includegraphics[width=\hsize]{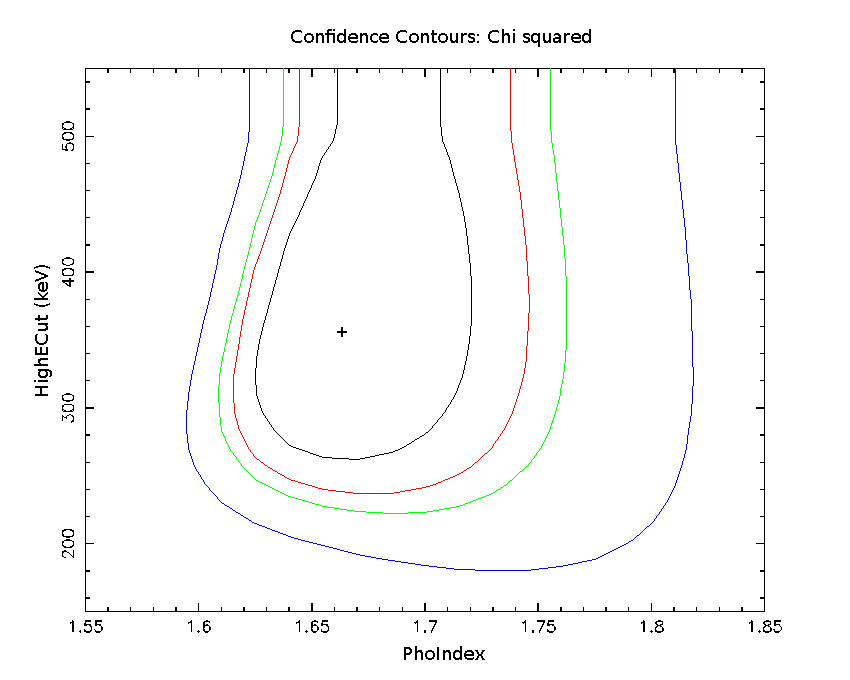}
\caption{Dependency of the $\chi^2$ on the photon index $\Gamma$ and cutoff energy $E_c$ of the cutoff powerlaw component for the \PCA\ in the cutoff powerlaw + log-parabola model. The cross represents the best fit model. The lines from the internal to the external show confidence contour at $\sigma =$ 1, 2, 2.706 (90\%), 5.}
\label{fig_parcont}
\end{figure}

In the \PCA\ the variation of the spectral index $\Gamma$ plays an important role: Fig. \ref{fig_parcont} shows the dependency of the $\chi^2$ on the photon index $\Gamma$ and the cutoff energy $E_c$. A reasonable good fit in the \PCA\ can be obtained only if $\Gamma \gtrsim 1.6$ . At $E_c = 360$ keV the minimum is located at $\Gamma = 1.66$ with $\chi^2_{red} = 1.02$, but the fit remains good for higher values of the photon index. The dependency on the cutoff energy is very weak: $\chi^2$ is almost constant for a large range of $E_c$ values. The data basically constrain only a lower limit to $E_c$.

\begin{table*}
\caption{Best fit parameters, reduced $\chi^2$ and degree of freedom for the \textit{broken powerlaw} model. The first column identifies the dataset. See text for details.}
\label{tab_bkp}
\centering
\begin{tabular}{ccccccc}
\hline\hline
\noalign{\smallskip}
Spectrum & $\Gamma_1$ & $E_b (MeV)$ & $\Gamma_2$ & $N_{bp}$ & $\chi_{red}^2$ & $d.o.f.$ \\
\noalign{\smallskip}
\hline
\noalign{\smallskip}
LAT f.s. & $1.55 \pm 0.02$ & $(7.8_{-1.3}^{+1.6})$ & $2.61 \pm 0.04$ & $(2.64_{-0.02}^{+0.03}) 10^{-2}$ & 1.376 & 40 \\
\noalign{\smallskip}
\hline
\noalign{\smallskip}
PCA f.s. & $1.65 \pm 0.02$ & $(1.6_{-0.7}^{+1.0})$ & $2.41_{-0.08}^{+0.09}$ & $(3.80 \pm 0.02) 10^{-2}$ & 1.987 & 34 \\
\noalign{\smallskip}
\hline
\noalign{\smallskip}
CGRO & $1.62 \pm 0.01$ & $(2.2_{-0.9}^{+2.9})$ & $2.43_{-0.15}^{+0.17}$ & $(0.91 \pm 0.03) 10^{-2}$ & 0.718  & 16 \\
\noalign{\smallskip}
\hline
\end{tabular}
\end{table*}

\begin{table*}
\caption{Best fit parameters, reduced $\chi^2$ and degree of freedom for the \textit{log-parabola} model. The first column identifies the dataset. See text for details.}
\label{tab_lp}
\centering
\begin{tabular}{ccccccc}
\hline\hline
\noalign{\smallskip}
Spectrum & $E_{0}$ (keV) & $\alpha$ & $\beta$ & $N_{lp}$ & $\chi_{red}^2$ & $d.o.f.$ \\
\noalign{\smallskip}
\hline
\noalign{\smallskip}
LAT f.s. & 1 & $1.24 \pm 0.03$ & $0.107 \pm 0.004$ & $(1.58_{-0.15}^{+0.17}) 10^{-2}$ & 2.311 & 41 \\
\noalign{\smallskip}
\hline
\noalign{\smallskip}
PCA f.s. & 1 & $1.41 \pm 0.02$ & $(9.74_{-0.33}^{+0.35})10^{-2}$ & $(2.90_{-0.19}^{+0.20}) 10^{-2}$ & 1.060 & 35 \\
\noalign{\smallskip}
\hline
\noalign{\smallskip}
CGRO & 1 & $1.38 \pm 0.03$ & $(9.81_{-0.59}^{+0.66})10^{-2}$ & $(0.70 \pm 0.03) 10^{-2}$ & 1.290 & 17 \\
\noalign{\smallskip}
\hline
\end{tabular}
\end{table*}

\begin{table*}
\caption{Best fit parameters, reduced $\chi^2$ and degree of freedom for the \textit{cutoff powerlaw + log-parabola} model. The first column identifies the dataset. See text for details.}
\label{tab_lpcop}
\centering
\begin{tabular}{cccccccccc}
\hline\hline
\noalign{\smallskip}
Spectrum & $\Gamma$ & $E_c$ (keV) & $N_{cp}$ & $E_{0}$ (keV) & $\alpha$ & $\beta$ & $N_{lp}$ & $\chi_{red}^2$ & $d.o.f.$ \\
\noalign{\smallskip}
\hline
\noalign{\smallskip}
LAT f.s. & 1.7 & 150 & 0.028 & 1 & $0.75_{-0.08}^{+0.06}$ & $0.16 \pm 0.01$ & $(2.0 \pm 0.4) 10^{-3}$ & 0.971 & 41 \\
\noalign{\smallskip}
\hline
\noalign{\smallskip}
PCA f.s. & 1.7 & 150 & 0.028 & 1 & $1.10_{-0.07}^{+0.05}$ & $0.13 \pm 0.01$ & $(6.3 \pm 1.0) 10^{-3}$ & 1.266 & 35 \\
\noalign{\smallskip}
\hline
\noalign{\smallskip}
CGRO & 1.7 & 150 & 0.009 & 1 & $0.61_{-0.21}^{+0.20}$ & $0.19 \pm 0.03$ & $(0.34_{-0.18}^{+0.38}) 10^{-3}$ & 0.098 & 16 \\
\noalign{\smallskip}
\hline
\end{tabular}
\end{table*}

\begin{table*}
\caption{Best fit parameters, reduced $\chi^2$ and degree of freedom for the \textit{cutoff powerlaw + log-parabola} model at LAT flaring time. Here the fit is done using PCA, JEM-X, ISGRI and SPI data. Parameters of the log-parabola are obtained fitting LAT data alone and used here as frozen parameters. See text for details.}
\label{tab_lpcop2}
\centering
\begin{tabular}{cccccccccc}
\hline\hline
\noalign{\smallskip}
$\Gamma$ & $E_c (keV)$ & $N_{cp}$ & $E_{0}$ (MeV) & $\alpha$ & $\beta$ & $N_{lp}$ & $\chi_{red}^2$ & $d.o.f.$ \\
\noalign{\smallskip}
\hline
\noalign{\smallskip}
$1.55 \pm 0.04$ & $(3.7_{-1.1}^{+3.0}) 10^2$ & $(2.52_{-0.26}^{+0.29}) 10^{-2}$ & 100 & 2.33 & 0.19 & $2.63~10^{-11}$ & 0.934 & 35 \\
\noalign{\smallskip}
\hline
\end{tabular}
\end{table*}

\begin{table*}
\caption{Best fit parameters, reduced $\chi^2$ and degree of freedom for the \textit{cutoff powerlaw + log-parabola} model.  using for the cutoff powerlaw the best fit parameters of Table \ref{tab_lpcop2}. Third line reports the best fit for the \PCA\ if $\Gamma$ is not frozen. The first column identifies the dataset. See text for details.}
\label{tab_lpcop3}
\centering
\begin{tabular}{cccccccccc}
\hline\hline
\noalign{\smallskip}
Spectrum & $\Gamma$ & $E_c$ (keV) & $N_{cp}$ & $E_{0}$ (keV) & $\alpha$ & $\beta$ & $N_{lp}$ & $\chi_{red}^2$ & $d.o.f.$ \\
\noalign{\smallskip}
\hline
\noalign{\smallskip}
LAT f.s. & 1.55 & 370 & 0.0252 & 1 & $0.39_{-0.08}^{+0.11}$ & $0.19 \pm 0.01$ & $(1.6_{-0.4}^{+0.9}) 10^{-4}$ & 0.859 & 41 \\
\noalign{\smallskip}
\hline
\noalign{\smallskip}
PCA f.s. & 1.55 & 370 & 0.0252 & 1 & $1.34_{-0.08}^{+0.20}$ & $0.08_{-0.03}^{+0.01}$ & $(3.5_{-1.4}^{+2.9}) 10^{-3}$ & 2.057 & 35 \\
\noalign{\smallskip}
\hline
\noalign{\smallskip}
PCA f.s. & $1.66_{-0.05}^{+0.09}$ & 370 & 0.0252 & 1 & $1.27_{-0.06}^{+0.08}$ & $0.11 \pm 0.01$ & $(8.0_{-1.2}^{+1.5}) 10^{-3}$ & 1.021 & 34 \\
\noalign{\smallskip}
\hline
\end{tabular}
\end{table*}

\section{Discussion}
\label{disc}

The analysis performed on the lightcurves and energy spectra of 3C 273 suggests different origin for the X-ray emission with respect to the $\gamma$-ray emission. The $\gamma$-ray emission at $E \gtrsim 100$ MeV is compatible with the relativistically beamed SSC coming from the jet. This blazar-like origin is suggested by the cross-correlation between radio and $\gamma$ rays. The X-ray component is compatible with thermal photon emitted by the accretion disk and scattered in the X-ray domain by inverse-Compton processes, possibly on a hot corona close to the central black hole, as inferred for Seyfert galaxies.


At a first attempt, the Seyfert-like component has been modelled with the identical cutoff-powerlaw component in the LAT and \PCA s, using the same parameters $\Gamma$, $E_c$ and $N_{cp}$ (Table \ref{tab_lpcop}). This means that we are able to explain the X-ray spectral variability with the variability of the blazar-like component only. However it is unlikely that the Seyfert-like component is not varying. In fact we can find slightly better fits using a different spectral index $\Gamma$ in the cutoff component (Table \ref{tab_lpcop3}). Furthermore, the \CGRO\ clearly shows that at the time of the multiwavelength campaign described in \citet{Lichti1995} the source was globally in a much lower state.

We note that \citet{Beaklini2014} reported a cross-correlation between 3C 273 Fermi-LAT data and radio data at 7 mm wavelength (43 GHz) claiming a delay between 120 and 170 days fully consistent with our estimate of $\sim$ 140 d (Sect. \ref{temp_conv}), and with similar conclusions on the link between radio and $\gamma$-rays. They explain the long-term variability of 3C 273 by variation of the Doppler factor in a 16-year jet-precessing model. The \LAT\ lasts roughly $\sim 100$ days, and according to the interpretation of \citet{Beaklini2014} it is at the time when the Doppler factor is highest. The \PCA\ spans a longer and not continuous time period: The simultaneous LAT data may thus be a composition of photons emitted with a noticeably different Doppler factor. Our data are thus broadly consistent with the precessing jet interpretation of these authors, although we cannot really test it with the chosen two different states.

\subsection{The convolution and the IC response}
\label{disc_conv}

Radio emission in radio loud AGN as 3C 273 is known to originate through synchrotron mechanism due to shocks in a relativistic jet \citep{Begelman1984,Marscher1985}. The response function used in the convolution (Sect. \ref{temp_conv}) hides the processes that happen in the propagation of the shock, like the cooling of the electrons and the IC mechanism. In the inner part of the jet the electrons are energetic enough to produce a flare in $\gamma$ rays through IC processes. Further out the flare is produced at longer wavelength because of the electron cooling and the expansion of the emitting region. 

The relative success of our convolution attempt suggests that the radio outbursts are closely related to the $\gamma$-ray activity. The effect of the convolution is to smear out the complex variability structure of the Fermi lightcurve, resulting in only two apparently distinct radio outbursts. Our convolution thus implies that the radio outbursts would actually be a blend of emission from electron accelerated in several shorter shocks.

It has to be pointed out however that longer flares in radio band may be the result of the superimposition of shorter flares: \citet{Valtaoja1999} decomposed radio lightcurves (22 and 37 GHz band) of an AGN sample into exponential flares. \citet{Hovatta2009} used this technique to derive the variability of Doppler boosting and Lorentz factor of the jets from the time scales of the flares. These works show that it is possible to decompose a long radio flare into several sequential components. VLBI data would allow us to identify the different radio components and then it is possible to look for direct correlation with $\gamma$-ray data. \citet{Jorstad2001} used this method to look for correlation between radio and $\gamma$-ray properties in a sample of blazars.

When building the synthetic radio lightcurve we had to vary the amplitude of the IC response from flare to flare (Table \ref{tab_pconv}). There are several physical parameters which can affect the IC process and may vary from shock to shock producing the observed variation in the amplitude of the IC response, but probably the most important are the size of the shocked region and the injected electron distribution. The bigger is the size of the shock or the more energetic are the injected electrons, the larger is the number of scattered photons, and hence the more prominent is the flare in $\gamma$ rays. Different knots of the jet can be ejected at slightly different angles with respect to the observer's line of sight or with a slightly different Lorentz bulk factor, resulting in a different Doppler factor \citep{Jorstad2001}. The emission of the knots with higher Doppler factor is hence boosted and shifted to higher frequencies, increasing consequently the total $\gamma$-ray emission. Both these considerations can explain the observed variation of the IC response amplitude.

\subsection{Interpretation of the log-parabola model}
\label{disc_logpar}

The energy spectrum of the LAT and \PCA s can be fitted with different parameter values. At first we note that the spectral index $\Gamma$ of the cutoff powerlaw and the spectral index $\alpha$ of the log-parabola are both harder in the \LAT. The best fit of the \LAT\ is obtained with $\Gamma = 1.55$ and $\alpha = 0.39$ while the best fit of the \PCA\ is with $\Gamma = 1.66$ and $\alpha = 1.27$ (first and third lines of Table \ref{tab_lpcop3}, respectively).

In the \PCA, the log-parabola model, because of the low statistics of the Fermi-LAT data, is sufficient to fit the dataset. Using the cutoff powerlaw + log-parabola model a degeneracy is added and the parameters can vary in a quite wide range.

The best fit yields a slightly steeper photon index $\Gamma$ of the cutoff component together with a lower photon index $\alpha$ of the log-parabola and a slightly increased curvature $\beta$. This means that with a softer cutoff powerlaw component the log-parabola index becomes harder in order to adjust the global slope of the spectrum (Table \ref{tab_lpcop3}). However exploring the range of the possible spectral parameter in the \PCA\ the $\beta$ parameter is always smaller and the $\alpha$ parameter is always larger with respect to the \LAT, meaning that the spectral variation between the two states is real. 
It suggests a harder spectrum with stronger curvature in the \LAT.

In order to find a possible physical interpretation of this difference we note that it is possible to show that a log-parabolic distribution of particles $n(\gamma)$ generates a log-parabolic photon spectrum using the $\delta$-assumption for the emission of a single particle. In this case the relation between the spectral indices is $\alpha = (s - 1)/2$ and $\beta = r/4$, where $s$ and $r$ are respectively the spectral index and the curvature in the log-parabolic particle distribution $n(\gamma)$.

Interpreting the log-parabola spectrum as the result of stochastic acceleration of the emitting electrons in a simple statistical approach, \citet{Tramacere2011} showed that $s \propto - \log ( \overline{\varepsilon} ) \left( \sigma_\varepsilon / \overline{\varepsilon}  \right)^{-2}$ (it can be obtained from their Eq. 9) and $r \propto \left(n \left(\sigma_\varepsilon / \overline{\varepsilon} \right)^2 \right)^{-1}$, where $n$ is the mean number of acceleration steps, and $\overline{\varepsilon}$ and $\sigma_\varepsilon^2$ are respectively the mean and the variance of the energy gain in each acceleration step. The spectral parameters are therefore related to the acceleration mechanism through the coefficient of variation $\sigma_\varepsilon / \overline{\varepsilon}$. The larger is $\sigma_\varepsilon / \overline{\varepsilon}$, the softer and less curved is the spectrum.

Our data show that the log-parabola component of the \LAT\ is harder and more curved than in the \PCA, and also that the peak energy of the SED is higher in the \LAT. In the stochastic acceleration interpretation, this means that the acceleration mechanism during the $\gamma$-ray flares is characterized by a lower coefficient of variation. This could result, for istance, because particles are uniformly boosted to higher energies through shocks.


\subsection{Synchrotron Self Compton variability}
\label{disc_ssc}

The SSC mechanism predicts constrains on the spectral variability which can be used to test the compatibility between the data and the model \citep[i.e. see][and reference therein for a detailed discussion on SSC]{Jones1974,Bloom1996}. \citet{Bloom1996} explored the consequences of the variations of the physical parameters of the emitting source (i.e. the magnetic field, the particle density distribution, the viewing angle) on the resulting spectrum in the frame of the SSC model.

We try to derive some informations on the physics of the X-rays and $\gamma$-rays emitting region in 3C 273 based on the differences between the LAT and \PCA s assuming that the blazar-like emission modelled as a log-parabola is due entirely to the SSC mechanism, without any External Compton contribute.
In order to do this we introduce first Eqs. \ref{eq_lpepeak} and \ref{eq_lpmax} which give the value of the log-parabola peak energy $E_p$ and the SED flux at the peak $E_p$ \citep{Massaro2004}:

\begin{equation}
\label{eq_lpepeak}
E_{p} = E_{0} ~ 10^{(2 - \alpha)/2\beta}
\end{equation}

\begin{equation}
\label{eq_lpmax}
\log \left(E_p F_{E}(E_p) \right) = \log N_{lp} + \frac{(2 - \alpha)^2}{4 \beta}.
\end{equation}

The peak energy can be used to estimate the variation of the Lorentz factor $\gamma$ of the emitting electrons. In the SSC the radiation field is produced by synchrotron emission, so that the energy released by a single electron is peaking at a frequency of $\displaystyle \nu_c \propto \gamma^2 \frac{eB}{m_e c}$. This radiation field then interacts with the electrons themself, and photons are scattered to higher energies through IC scattering, increasing their frequency by another factor $\gamma^2$.

The ratio between the energy peaks of LAT and \PCA s can provide information about the variation of $\gamma$ in the two spectral states. This ratio can be written as

\begin{equation}
\label{eq_epeakrat}
\frac{E_L^C}{E_P^C} \sim \frac{\gamma_L^2 E_L^S}{\gamma_P^2 E_P^S} \sim \left( \frac{\gamma_L}{\gamma_P} \right)^4 \frac{B_L}{B_P},
\end{equation}

\noindent where the subscripts L and P stand for LAT and \PCA s, respectively, and the superscripts C and S stand for IC and synchrotron. Taking the energy peak of the log-parabola as $E^C$ and hence using Eq. \ref{eq_lpepeak} and the best fit parameters of the spectral states in Table \ref{tab_lpcop} we get $\gamma_L / \gamma_P \simeq 1.6$ if we assume $E_L^S= E_P^S$ or $\gamma_L / \gamma_P \simeq 1.2$ if we assume $B_L = B_P$. Particles would thus be accelerated to higher energies during the \LAT.\\


To further investigate the physical parameters of the emitting region we can use the flux at the peak. We first rewrite the peak energy of the IC spectra in unit of $\displaystyle \eps = \frac{h\nu}{m_e c^2}$ as

\begin{equation}
\label{eq_epeak}
\eps_p = \frac{\delta_D}{1 + \mathrm{z}}\gamma^4 \varepsilon_B
\end{equation}

\noindent where $\displaystyle \varepsilon_B = \frac{\hbar eB}{m_e^2 c^3}$ and $\delta_D$ is the Doppler factor. The $\gamma$ factor here is that of the electrons emitting at the peak energy.

This photon energy is used in Eq. \ref{eq_flssc}, which gives the observed SSC $\nu F_\nu$ spectrum \citep[Ch. 7, eq. 7.97]{Dermerbook}

\begin{equation}
\label{eq_flssc}
\begin{array}{rcl}
\displaystyle E F_{E}(\eps) & = & \displaystyle \delta_D^4 \frac{c\sigma_T^2 R' U_B V'}{12 \pi d_L^2} \left( \frac{\eps}{\varepsilon_B} \right)^{3/2} \times\\
& & \displaystyle \int_0^{\min(\eps', 1/\eps')} d\eps_i' \eps_i'^{-1} n_e' \left( \sqrt{\frac{\eps'}{\eps_i'}} \right) n_e' \left( \sqrt{\frac{\eps_i'}{\varepsilon_B}} \right).
\end{array}
\end{equation}

Equation \ref{eq_flssc} is obtained in the $\delta$-function approximation and limiting the IC interation to the Thompson regime (upper limit of the integral). Prime quantities are measured in the rest frame of the source. $R'$ is the size of the source ($V' \sim R'^3$ being the volume), $\displaystyle U_B = \frac{B^2}{8\pi}$ is the energy density of the magnetic field. The electron distribution $n_e(\gamma)$ appears twice, the former to account for the IC scattering, the latter for the synchrotron emission. $\eps_i'$ is the energy of the synchrotron photons which are scattered.

Equations \ref{eq_epeak} and \ref{eq_flssc} relate the peak energy $E_p$ and the $\nu F_\nu$ spectrum to the physical quantities of the source. Varying the physical parameters we can explore how the peak energy and the $\nu F_\nu$ flux calculated at the peak energy vary thereby and compare these trends with the data.

According to Eq. \ref{eq_epeak} the observed peak energy depends on the Lorentz factor $\gamma$ of the electrons at the peak and the magnetic field $B$ (we do not consider here any variation of the Doppler factor $\delta_D$ of the emitting region). The value of the observed $E_p$ in LAT and \PCA s is reproduced by fixing arbitrary one of the two parameters and setting properly the remaining one. Then we can compute the SSC $\nu F_\nu$ flux at the peak energy with Eq. \ref{eq_flssc}, provided that the remaining parameters are fixed.

To solve Eq. \ref{eq_flssc} we assume a log-parabolic electron distribution, for which we can derive the parameters $s$ and $r$ from the spectral fit as mentioned in Sect. \ref{disc_logpar}, and thus solve the integral numerically. We used the parameters of the best fit from Table \ref{tab_lpcop} for the LAT and \PCA s to derive the parameters of the electron distribution.

\begin{figure}
\centering
\includegraphics[angle=-90, width=\hsize]{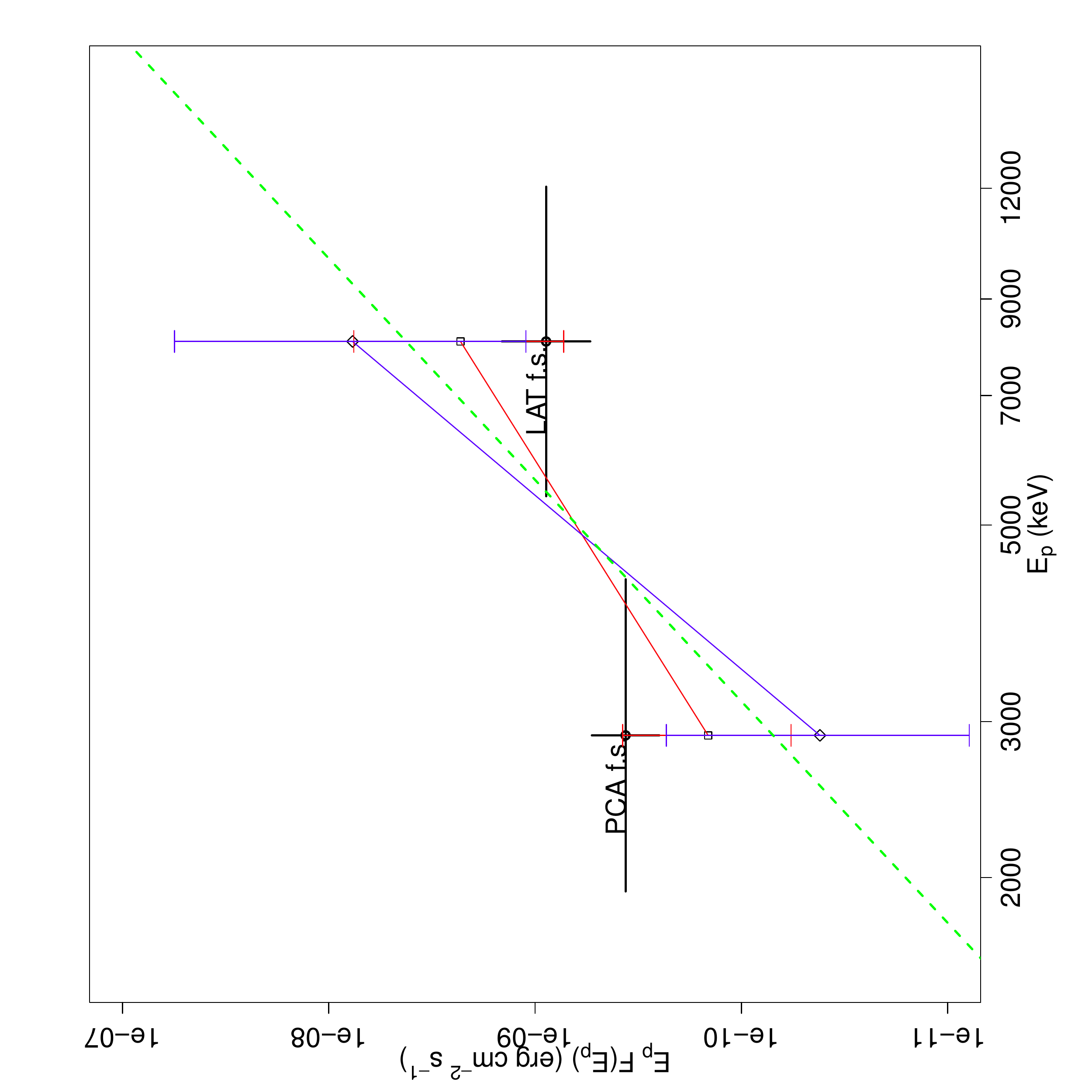}
\caption{Trend of variation of the SSC flux in the $E_p - E_p F_E (E_p)$ plane. The black points correspond to the LAT and \PCA s. The red line connects the two points with different Lorentz factor $\gamma$, the blue line connects the two points with different magnetic field $B$. The parameters are listed in Table \ref{tab_varypar}. The dotted green line shows the trend associated to $\delta_D$ variation. See text for details.}
\label{fig_varypar}
\end{figure}

Figure \ref{fig_varypar} shows these points together with the observations of the LAT and \PCA s, Table \ref{tab_varypar} list the exact energy $E_p$ and associated SED flux, together with the corresponding values of the magnetic field $B$ and Lorentz factor $\gamma$.\\

\begin{table}
\caption{List of the varying parameters $\gamma$ and $B$ used to draw the points in Fig. \ref{fig_varypar}. The remaining parameters are fixed to the following values: $\delta_D = 9$, $R' = 1.6 \; 10^{15}$ cm, $\displaystyle n_e' = 2.7 \; 10^{6} \; \mathrm{electrons}/\mathrm{cm}^3$, z = 0.158, $d_L = 2.30 \; 10^{27}$ cm, estimated for the redshift of 3C 273 in a flat universe ($\Omega_\Lambda = 1 - \Omega_M$) with $H_0$ = 71 km/s Mpc and $\Omega_M = 0.27$.}
\label{tab_varypar}
\centering
\begin{tabular}{cccc}
\hline\hline
\noalign{\smallskip}
$B$ (Gauss) & $\gamma$ & $E_p$ (keV) & $E_p F_E(E_p) \left( \frac{\mathrm{erg}}{\mathrm{cm}^2 s} \right)$ \\
\noalign{\smallskip}
\hline
\noalign{\smallskip}
1.4 & $2.8 \; 10^3$ & $8.0 \; 10^3$ & $2.4_{-1.7}^{+5.1} \; 10^{-9}$ \\
\noalign{\smallskip}
\hline
\noalign{\smallskip}
1.4 & $2.2 \; 10^3$ & $2.9 \; 10^3$ & $1.5_{-0.9}^{+2.3} \; 10^{-10}$ \\
\noalign{\smallskip}
\hline
\noalign{\smallskip}
2.3 & $2.5 \; 10^3$ & $8.0 \; 10^3$ & $7.9_{-6.8}^{+48.0} \; 10^{-9}$ \\
\noalign{\smallskip}
\hline
\noalign{\smallskip}
0.8 & $2.5 \; 10^3$ & $2.9 \; 10^3$ & $3.1_{-2.3}^{+20.0} \; 10^{-11}$ \\
\noalign{\smallskip}
\hline
\end{tabular}
\end{table}

We choose to investigate variation in $B$ and $\gamma$ because in our simple description, $E_p$ depends on these two physical parameters only. $\delta_D$ and $R'$ have been fixed to $\delta_D = 9$ \citep{Jorstad2005} and $R' = 1.6 \; 10^{15}$ cm \citep{Rani2013}, the remaining parameter $n_e'$ has been calculated to reproduce the observed flux. The total electron density $n_e'$ is the same in both states, i.e. the integral of the log-parabolic distribution $n_e'(\gamma)$ is forced to be constant adjusting the normalization constant. The electron density value needed to reproduce the observed flux is $n_e' = 2.7 \; 10^{6} \; \mathrm{electrons}/\mathrm{cm}^3$. This looks like a big value, but is low enough for an optically thin emitting region: the optical depth $\tau$ for inverse-Compton scattering in Thompson regime can be estimated to be $\tau \sim R' \sigma_T n_e' \sim 10^{-3}$.

Figure \ref{fig_varypar} shows that the observed variability between the LAT and \PCA s is better described by a variable Lorentz factor $\gamma$ with a constant magnetic field B, rather than a constant $\gamma$ and a variable B. The confidence intervals are obtained accounting for the errors on the spectral parameters: the parameters are modified by their errors at $1 \sigma$ and then the flux is recalculated with Eq. \ref{eq_flssc}. This indicates that the variation of the Lorentz factor $\gamma$ of the electrons is likely responsible for the difference between the LAT and the \PCA s. We note that variations of $R'$ or $n_e'$ do not produce a variation in $E_p$.

We investigated also if the variation could be explained by the variation of the Doppler factor $\delta_D$ of the emitting region. Different knots may have different Doppler factors because they may be emitted at a different angle or with different velocity, as assumed in the varying $\delta_D$-factor model of \citet{Beaklini2014}. To test this, we used the same electron distribution $n_e'(\gamma)$ for the LAT and \PCA s assuming that the only varying parameter between the two states is $\delta_D$. In this case $E_p$ increases linearly with $\delta_D$, while $E_p F_E(E_p)$ increases as $\delta_D^4$ (see Eq. \ref{eq_flssc}, the trend shown by the dotted green line in Fig. \ref{fig_varypar}). This trend is not compatible with the data.

In conclusion, a variation of the electron distribution and consequently a variation of the electron Lorentz factor $\gamma$ corresponding to the peak of the electron energy distribution seems to be the most likely driver for the difference between the LAT and \PCA s.

\subsection{Location of the $\gamma$-ray emitting region}
\label{disc_locgamma}
The spectrum observed by Fermi-LAT is not significantly absorbed up to $\sim 3$ GeV. Photons below 3 GeV could produce pairs by interacting with target photons of energies above ~ 0.3 keV. The latter are probably emitted from the accretion disk and corona within 100 gravitational radii from the central black hole. The $\gamma$-ray emitting region should be located outside of that region, which is not very constraining.

Since the sensitivity of Cherenkov telescopes is not sufficient to detect 3C 273 in the TeV domain, where the pair absorption signature should be present if the $\gamma$-ray emitting region is inside the broad line region, the position of the $\gamma$-ray source cannot be constrained with respect to the size of the broad line region \citep{Donea2003}.

\section{Summary and Conclusions}
\label{concl}
The high energy spectral variability of 3C 273 has been studied from $\sim 1$ keV up to $\sim 10$ GeV and compared with radio data. We presented Fermi-LAT, RXTE-PCA and Radio (37 GHz) lightcurves and multiwavelength spectra of RXTE-PCA, Fermi-LAT and JEMX, ISGRI, SPI on board INTEGRAL, built for two different spectral states, namely the \LAT, i.e. when the source is flaring in Fermi-LAT, and the \PCA, i.e. when the source is flaring in RXTE-PCA.

Timing analysis (Sect. \ref{temp}) indicates the presence of two different components in the considered energy band. Gamma ray data correlate with radio data, but not with X-ray data, confirming that the emissions in $\gamma$ rays and radio have a common origin, while the origin of the X-ray emission is different. The convolution between the $\gamma$-ray lightcurve and the SSC response function reproduces the radio signal (Sect. \ref{temp_conv}). This indicates that the data are compatible with the synchrotron mechanism in the frame of a shock-in-jet scenario \citep{Marscher1985,Turler1999,Turler2000}. The necessity to shrink the response function obtained by low energy data suggests that what is seen as a single flare in the radio lightcurve may be in this case the blended emission from several shorter shocks, which have been detected separately in the $\gamma$-ray lightcurve.

Spectral fitting of the high energy data (Sect. \ref{spec}) shows that a single model emission as a cutoff powerlaw or a log-parabola cannot fit adequately the whole SED from $\sim 1$ keV up to $\sim 10$ GeV. Knowing from timing analysis that X-rays and $\gamma$ rays have different origin, we used a two components spectral model: a cutoff powerlaw at low energies and a log-parabola at high energies, the former modelling the Seyfert-like emission, the latter modelling the jet emission. We found that for the \LAT\ the powerlaw and log-parabola components are harder than for the \PCA, and the \LAT\ spectrum has a stronger curvature at the SED peak. The spectral variability between the two states can be understood if it is driven by a change in the electron distribution with a peak energy increasing by $\sim 30 \%$ during the \LAT.\\

Unlike typical blazars, the high energy spectrum of 3C 273 is peaked in the MeV energy range, below the Fermi-LAT lower energy threshold. Even during the flare, the peak of the emission remains undetected. As there are no experiments sensitive enough in the MeV, it is difficult to study in detail the spectral properties of the $\gamma$-ray emission of 3C 273. Nevertheless, the shape of the $\gamma$-ray spectrum of 3C 273 is very similar to the shape of the GeV peaked blazar detected by Fermi, which are usually fitted with the log-parabola model, suggesting that the electron distribution looks universal in very different conditions.

\begin{acknowledgements}
This work is based on observation with INTEGRAL, an ESA project with instrument and science data center funded by ESA member states (especially the PI countries: Denmark, France, Germany, Italy, Switzerland, Spain), Czech Republic and Poland, and with the participation of Russia and the USA. The SPI project has been completed under the responsibility and leadership of CNES/France. The \Metsa\ team acknowledges the support from the Academy of Finland to our observing projects (numbers 212656, 210338, 121148, and others).
\end{acknowledgements}

\end{document}